\documentclass[12pt]{article} 

\usepackage{graphicx, url, algorithm2e}
\usepackage{amssymb, amsmath, amsthm, color, xcolor, nicefrac,pgfplots}

\usepackage{fullpage}

\newtheorem{theorem}{Theorem}
\newtheorem{lemma}[theorem]{Lemma}
\newtheorem{definition}[theorem]{Definition}

\newcommand {\Set} [1] {\left\{ #1 \right\}}
\newcommand {\Angle} [1] {\left\langle #1 \right\rangle}
\DeclareMathOperator*{\argmin}{arg\,min}
\DeclareMathOperator {\cost} {cost}

\newcommand{\given}{\mid}
\newcommand{\poly}{\mathrm{poly}}
\newcommand {\E} {\mathbb{E}}
\newcommand {\calZ} {{\cal{Z}}}
\newcommand {\calW} {{\cal{W}}}

\newcommand {\bbR} {\mathbb{R}}

\newcommand {\calC} {{\cal{C}}}
\newcommand {\calD} {{\cal{D}}}
\newcommand {\calE} {{\cal{E}}}

\begin{document}
\title{A bi-criteria approximation algorithm for $k$ Means}

\author{Konstantin Makarychev \\ Microsoft Research \and
Yury Makarychev \\ Toyota Technological Institute at Chicago \and
Maxim Sviridenko \\ Yahoo Labs \and
Justin Ward\footnote{Work supported by EPSRC grant EP/J021814/1.} \\ University of Warwick}

\maketitle

\begin{abstract}
We consider the classical $k$-means clustering problem in the setting bi-criteria approximation, in which an algoithm is allowed to output $\beta k > k$ clusters, and must produce a clustering with cost at most $\alpha$ times the to the cost of the optimal set of $k$ clusters.  We argue that this approach is natural in many settings, for which the exact number of clusters is a priori unknown, or unimportant up to a constant factor.
We give new bi-criteria approximation algorithms, based on linear programming and local search, respectively, which attain a guarantee $\alpha(\beta)$ depending on the number $\beta k$ of clusters that may be opened.  Our gurantee $\alpha(\beta)$ is always at most $9 + \epsilon$ and improves rapidly with $\beta$ (for example: $\alpha(2)<2.59$, and $\alpha(3) < 1.4$).  Moreover, our algorithms have only polynomial dependence on the dimension of the input data, and so are applicable in high-dimensional settings.
\end{abstract}

\section{Introduction}

The $k$-means clustering problem is one of the most popular models for unsupervised Machine Learning. The problem is formally defined as follows.
\begin{definition}
In the \emph{$k$-means} problem, we are given a set $X$ of $n$ points $x_1,\dots, x_n$ in ${\mathbb R}^p$ and an integer parameter $k \geq 1$. Our goal is to partition $X$ into $k$
clusters $S_1,\dots, S_k$ and assign each cluster a center $a_i$ so as to minimize the cost $\sum_{i=1}^k \sum_{x_j\in S_i} \|x_j - a_i\|^2$.
\end{definition}

The most common heuristic for $k$-means is Lloyd's algorithm introduced in 1957 \cite{Lloyd, LloydJournal}.
Llloyd's algorithm starts with some initial solution and then iteratively improves it by alternating two steps: at the first step, the algorithm
picks the optimal clustering for the current set of centers; at the second step, the algorithm picks the optimal set of centers for the current
clustering.  While we know that the algorithm performs well on well-clusterable data \cite{OR} it performs arbitrarily badly on general instances. There exist many variants
of this algorithm and many heuristics for picking the initial solution.
Unfortunately, none of them give a constant (not depending on $k$)
factor approximation.  One of the most popular ones is the $k$-means{\sc++} algorithm that has an $O(\log k)$-approximation factor \cite{AV}.

The general $k$-means clustering problem has recently been shown to be APX-hard, ruling out a PTAS in the general case \cite{apx-hard}.  However, a variety of PTASes exist for special cases of the problem.  Inaba, Katoh, and Imai \cite{Inaba} gave a $(1+\varepsilon)$-approximation algorithm for the case in which the number of clusters, $k$, and the dimension of the space, $d$, are fixed. Since then many more PTASes were proposed for other special cases.  Additionally, there are many results showing the NP-hardness for several special cases \cite{hard1,hard2,hard3}.

The best constant factor approximation algorithm for the general case of the problem was proposed by \cite{Kanungo2004} a decade ago.
Their algorithm gives $9+\varepsilon$ factor approximation. Using the connection with $k$-median problem \cite{approx1} designed an alternative constant factor approximation algorithm.  \cite{approx2} showed that running the $k$-means{\sc++} algorithm for more steps gives an $\alpha = 4+\varepsilon$ factor approximation by opening $\lceil 16(k + \sqrt{k}) \rceil$ centers, and also showed how to modify the resulting solution to obtain a set of $k$ centers attaining an $O(1)$ factor guarantee.

In most practical applications the target number $k$ of clusters is not fixed in advance, rather we would like to find a number $k$ that provides a well-clusterable solution.  Here, we show how to substantially improve the approximation factor by slightly violating the constraint on the number of clusters.
We present bi-criteria approximation algorithms for the general case of the problem. A $(\beta, \alpha)$  bi-criteria approximation algorithm
finds a solution with $\beta k$ clusters, whose cost is at most $\alpha$ times the optimal cost of a solution using $k$ clusters.  In contrast to the approach of \cite{approx2}, our algorithms find an approximate solution for every $\beta>1$. Our approximation is always at most $9$, and decreases rapidly with $\beta$.  In particular, we obtain a $4$-approximation by opening only $1.65 k$ centers, improving over previous results \cite{approx2} by a factor of nearly 10, and obtain improved approximation factors $\alpha(\beta)$ as $\beta$ continues to grow.  For example, $\alpha(1.3)<6.45$, $\alpha(1.5)<4.8$; $\alpha(2)<2.59$, and $\alpha(3) < 1.4$.
In general, we argue that in many applications the number of clusters is not important as long as it approximately equals $k$. For these applications we can obtain an approximation factor very close to 1.
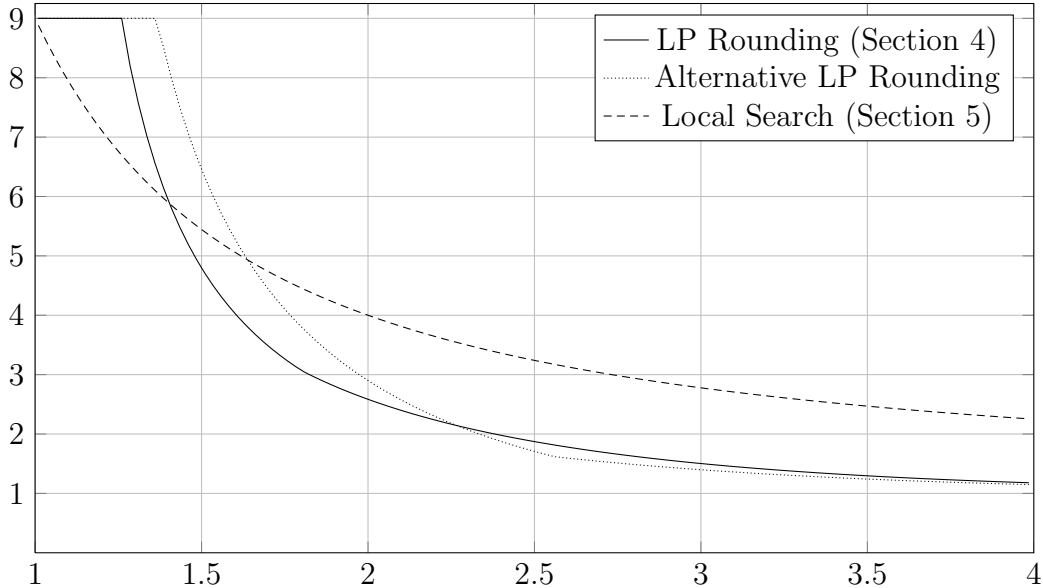
\begin{figure}
\centering
\begin{tikzpicture}
\begin{axis}[width=0.9\textwidth,
height=3.5in,
ymax=9.25, ymin=0, ytick={1,2,3,4,5,6,7,8,9},
xmin=1, xmax=4, xtick={1,1.5,2,2.5,3,3.5,4},
grid=major,
legend entries={LP Rounding (Section \ref{sec:appr-k-medi}),Alternative LP Rounding, Local Search (Section \ref{sec:local-search})},
]
col sep=comma]{plots.csv};
\addplot[color=black]table [x expr=\thisrowno{0},
y expr=\thisrowno{2},
col sep=comma]{plots.csv};
\addplot[color=black,densely dotted]table [x expr=\thisrowno{0},
y expr=\thisrowno{3},
col sep=comma]{plots.csv};
\addplot[color=black,densely dashed]table [x expr=\thisrowno{0},
y expr=\thisrowno{4},
col sep=comma]{plots.csv};
\end{axis}[]
\end{tikzpicture}
  \caption{Approximation ratios obtained from opening $\beta k$ centers}\label{fig:performance}
\end{figure}


We give three bi-criteria algorithms---two based on linear programming and one based on local search.  We show the algorithms' approximation factors as a function of $\beta$ in Figure~\ref{fig:performance}.  Note that our linear programming algorithm attains a better approximation $\alpha$ for large $\beta$, while the local search algorithm is better for $\beta$ near 1.

Both of our algorithms are based on a reduction from the $k$-means problem, in which cluster centers may be placed at any point in ${\mathbb R}^p,$ to the following $k$-median problem, in which we are restricted to a given, discrete set of candidate cluster centers with specified distances from each point.
As part of reduction, we utilize dimensionality reduction to ensure that the number of discrete candidate centers that must be considered is polynomial
in both the number of points $n$ and in the dimension $p$.
\begin{definition}
In the \emph{$k$-median problem}, we are given a set of points $\cal D$, a set of potential center locations ${\cal C}$ and a distance function
$(d(i,j))_{i \in \calC, j\in \calD}$.
The cost of assigning point $j$ to center $i$ is $d(i,j)$.
Our goal is to open at most $k$ centers and assign each point to a center so as to minimize the total cost.
\end{definition}
The first approximation algorithms for the $k$-median problem were given by \cite{LinVitter}, who gave an LP-rounding algorithm that attains an approximation factor of $1 + \varepsilon$ by opening $O(k\ln n)$ centers (i.e. a $(1+\epsilon, O(\ln n))$ bi-criteria approximation).  In further work, \cite{LinVitter2} showed that if the distance function $d$ is a \emph{metric}, it is possible to obtain a $2(1 + \varepsilon)$ approximation algorithm by opening only $(1 + 1/\varepsilon)k$ centers. The first constant-factor approximation for the metric $k$-median problem using only $k$ centers was obtained by \cite{Arya2004}, who showed that a simple local search algorithm gives a $3 + \varepsilon$ approximation.  This remained the state of the art, until recently, when \cite{LiSvensson} gave a $2.732 + \varepsilon$ approximation algorithm based on LP rounding.  Subsequently, this has been  improved to $2.611 + \varepsilon$ by \cite{Byrka}.

Unfortunately, our resulting $k$-median instance is non-metric, and so we must employ an alternative to the standard triangle inequality in our analysis.  In the case of our LP-based algorithms, we use the fact that our reduction produces instances satisfying a 3-relaxed 3-hop triangle inequality, a concept that we define in Section \ref{sec:preliminaries}.  In the case of local search, we note that given any partition of points of $\mathbb{R}^p$ into clusters $S_1,\ldots,S_k$, the optimal location of each $k$-means cluster $S_i$'s center is the centroid of all points in $S_i$.  This, combined with the fact that our reduction to $k$-median approximately preserves the $k$-means cluster costs allows us to employ a similar approach to that of \cite{Kanungo2004}.

\subsection{Our Results}
We give three approximation algorithms. The first algorithm is based on linear programming. It gives
$$
\alpha_1(\beta) = 1+e^{-\beta}\Big(\frac{6\beta}{1-\beta} + \frac{(\beta-1)^2}{\beta}\Big)
$$
approximation (see also~(\ref{eq:exact-LP-bound}) for a slightly tighter bound). The second algorithm is based on local search. It gives
$$
\alpha_2 (\beta) = (1+O(\varepsilon)) \Big(1+\frac{2}{\beta}\Big)^2
$$
approximation. The third algorithm is also based on linear programming. It gives
$$\alpha_3(\beta)=\max \big(1+8e^{-\beta}, \frac{\beta (e^{-1} + 8e^{-\beta})}{\beta -1}\big)$$
approximation. The algorithm is similar to the first algorithm, but it uses 
pipage rounding (see~\cite{AS}) instead of randomized rounding. In the conference version of the paper, we omit
the description of the third algorithm. The approximation factors are shown in Figure~\ref{fig:performance}.

In Section \ref{sec:preliminaries} we introduce the notation that we shall use throughout the rest of the paper and review standard notions related to both the $k$-means and  $k$-median problems.  In Section \ref{sec:reduction-from-k}, we give the details of our reduction to $k$-median.  Finally, in Sections \ref{sec:appr-k-medi}  and \ref{sec:local-search}, respectively, we present our main LP-based algorithm and  local search algorithm for the resulting $k$-median instances.


\section{Preliminaries}
\label{sec:preliminaries}

We now fix some notation, and recall some basic properties of $k$-means solutions and the standard linear program for the $k$-median problem.  Additionally, we define the notion of an $\alpha$-relaxed 3-hop triangle inequality, which will be crucial to the analysis of our LP-rounding algorithms.

\subsection{$k$-means}
Consider a given instance of the $k$-means problem, specified by a set of points $X \in \bbR^p$.  Given a partition $S = \Angle{S_1,\ldots,S_k}$ of $X$ and a set $C = \Angle{c_1,\ldots,c_k}$ of centers in $\bbR^p$, denote by $\cost_X(S,C)$ the total cost of the clustering that, for each $1 \le i \le k$ assigns each point of $S_i$ to the center $c_i$:
\begin{equation*}
\cost_X(S,C) = \sum_{i = 1}^k \sum_{x \in S_i}\|x - c_i \|^2.
\end{equation*}
Note that to describe an optimal solution to the $k$-means problem, it is sufficient to specify either all clusters or all centers in the solution.  Indeed, given a list of clusters $S_1, \dots, S_k$, we can find the optimal assignment of centers $c_i$ for it: the optimal choice of center $c_i$ for $S_i$ is $\frac{1}{|S_i|}\sum_{x_j \in S_i} x_j$.  For this choice of $c_i$, we have
\begin{equation}
\label{eq:costS}
\sum_{x\in S_i} \|x - c_i\|^2 = \frac{1}{2|S_i|} \sum_{x', x''\in S_i} \|x' - x''\|^2.
\end{equation}
Given a partition $S = \Angle{S_1,\ldots,S_k}$ of $X$ into clusters, we then denote by $\cost_X(S)$ the cost of this optimal choice of centers.  That is,
\begin{equation*}
\cost_X(S) =  \sum_{i=1}^k  \frac{1}{2|S_i|} \sum_{x', x''\in S_i} \|x' - x''\|^2.
\end{equation*}

Similarly, given a list $C$ of centers $c_1,\dots,c_k$, we can find the optimal partition $S = \Angle{S_1, \ldots, S_k}$ of $X$ into clusters.  For each $c \in C$, let $N_C(c)$ be the set of those points $x\in X$ that are closer
to $c_i$ than to other centers $c_j \neq c_i$ (if a point $x$ is at the same distance from several centers, we break the ties arbitrarily).   The optimal partition for $C$ then sets $S_i = N_C(c_i)$.  Given a set $C$ of $k$ centers, we define $\cost_X(C) \equiv \cost_X(T)$, where $T=\Angle{N_C(c_1),\ldots,N_C(c_k)}$ is the partition induced by $C$.

\subsection{$k$-median}
We will reduce a given instance $X$ of $k$-means problem to an instance of the $k$-median problem, specified by $\langle \calD, \calC, d \rangle$.  By analogy with $k$-means problem, we can consider a partition $S = S_1,\ldots,S_k$ of demand points from $\calD$, and then consider the best choice of a single center for each partition.  We denote the cost of this choice by $\cost_{\calD, d}(S)$:
\[
\cost_{\calD,d}(S) = \sum_{i = 1}^k \min_{x \in \calC} \sum_{j \in S_i} d(x,j).
\]
Similarly, given a list of $k$ centers $C = \Angle{c_1,\ldots,c_k}$, let $N_C(c_i)$ be the set of those demand points $x \in \calD$ that are closer (according to the distance function $d$) to $c_i$ than to any other center in $C$ (again, if a point $x$ is at the same distance from several facilities, we break ties arbitrarily).  As in the case of $k$-means, we define $\cost_{\calD, d}(C) \equiv \cost_{\calD, d}(T)$ where $T = \Angle{N_C(c_1),\ldots,N_C(c_k)}$ is the partition of $\calD$ induced by $C$.

Although the distance function $d$ in our $k$-median instances will not satisfy the standard triangle inequality, we can show that it satisfies a relaxed variant of the following sort:
\begin{definition}
We say that $d$ satisfies an \emph{$\alpha$-relaxed $3$-hop triangle inequality} on $\calD \cup \calC$ if, for any
$j,j'\in \calD $ and $i,i'\in \calC$, we have
$$d(i,j)\le \alpha\left(d(i,j)+d(i',j')+d(i',j)\right).$$
\end{definition}
Specifically, we shall show that the distances produced by our reduction satisfy a $3$-relaxed $3$-hop triangle inequality.

\section{Reduction from $k$-means to $k$-median}
\label{sec:reduction-from-k}
We now give the details of our reduction from the $k$-means to the $k$-median problem.
In the $k$-median problem, a finite set $\calC$ of candidate centers is specified, while in the $k$-means problem, the ideal center for each cluster $S_i$ of points is given by the centroid of $S_i$.  Ideally, we want to ensure that for every possible centroid of the original $k$-means instance, there is some nearby candidate centers in $\calC$.  The following notion of an $\varepsilon$-approximate centroid set, introduced by \cite{Mat}, captures this requirement.
\begin{definition}\label{def:centroid}
A set of points $\calC\subset \bbR^p$ is an $\varepsilon$-approximate centroid set for $X\subset \bbR^p$
if for every  $S\subset X$,
$$\min_{c\in \calC} \sum_{x\in S}\|x-c\|^2 \leq (1+\varepsilon) \min_{c\in \bbR^p} \sum_{x\in S}\|x-c\|^2.$$
\end{definition}
Observe that if $\calC$ is an $\varepsilon$-approximate centroid set for $X$, then for every set of
$k$ centers $C$ (in particular, for the optimal set $C^*$), there exists a $k$-point subset
$\widetilde{C}\subset \calC$ such that
$$\cost_X({\widetilde{C}}) = \sum_{x\in X} \min_{c\in \widetilde{C}} \|x - c\|^2 \leq  (1+\varepsilon) \sum_{x\in X} \min_{c\in C} \|x - c\|^2 =  (1+\varepsilon)\cost_X(C).$$
Thus, if we restrict our search for $k$ center points in $k$-means problem to only those points of $\calC$, we lose at most a factor of $(1+\varepsilon)$.

Matou\v{s}ek showed that for every set $X$ in $\bbR^p$ and $\varepsilon >0$, there exists an $\varepsilon$-approximate centroid set of size $O(|X| \varepsilon^{-p}\log (1/\varepsilon))$.
\begin{theorem}[Theorem 4.4 in \cite{Mat}]\label{thm:matousek}
Given an $n$-point set $X\subset \bbR^p$ and $\varepsilon >0$, an $\varepsilon$-approximate centroid set for $X$
of size $O(n \varepsilon^{-p}\log (1/\varepsilon))$ can be computed in time
$O(n \log n + n \varepsilon^{-p}\log(1/\varepsilon))$.
\end{theorem}
Unfortunately, in our setting, the dimension $p$ of the space in which points $x_1,\dots, x_n$ lie may be as large as $n$.  Thus, in order to apply Theorem \ref{thm:matousek}, we first embed $X$ into a low-dimensional space using the
 Johnson--Lindenstrauss transform.
\begin{theorem}[Johnson--Lindenstrauss Flattening Lemma]\label{thm:JL}
For every set of points $X$ in $\bbR^p$ and $\varepsilon \in (0,1)$,
there exists a map $\varphi$ of $X$
into $\tilde{p} = O(\log |X|/\varepsilon^2)$ dimensional space such that
\begin{equation}\label{eq:bi-lip}
\|x - y\|_2^2 \leq \|\varphi(x) - \varphi (y)\|_2^2 \leq (1+\varepsilon)\|x - y\|_2^2.
\end{equation}
We say that the map $\varphi$ is a \emph{dimension reduction transform} for $X$.
\end{theorem}
Given an instance $X$ of $k$-means, we apply the dimension reduction transform to $X$, get a set $X'\subset \bbR^{\tilde p}$, and then find an $\varepsilon$-approximate centroid set $\cal C$ to $X'$. We obtain an instance $\langle X', \calC, d\rangle$ of $k$-median with the squared Euclidean distance $d$. We show in Theorem~\ref{thm:reduction-main} that the value of this instance is within a factor of $(1+\varepsilon)$ of the value of instance $X$ of $k$-means, and, moreover, that there is
a one-to-one correspondence between solutions of instance $\langle X', \calC, d\rangle$ and solutions of instance $X$. We prove Theorem~\ref{thm:reduction-main} in
 Section~\ref{sec:reduction-proofs}.
\begin{theorem}\label{thm:reduction-main}
1. For every $\varepsilon \in (0,1/2)$, there  exists a polynomial-time reduction from $k$-means to $k$-median with
distance function that satisfies the 3-relaxed 3-hop triangle inequality. Specifically, given an instance $X$ of $k$-means,
the reduction outputs an instance $\langle \calD, \calC, d\rangle$ of $k$-median with
$|\calD| = |X|$, $|\calC| = n^{O(\log(1/\varepsilon)/\varepsilon^2)}$, and distance $d$ that satisfies the 3-relaxed 3-hop triangle inequality such that
$$\mathrm{OPT}_X \leq \mathrm{OPT}_{\langle \calD, d\rangle} \leq (1+\varepsilon) \mathrm{OPT}_X,$$
where $\mathrm{OPT}_X $ is the value of the optimal solution to $X$ and $\mathrm{OPT}_{\langle \calD,  d\rangle}$ is the value of the optimal solution
to $\langle \calD, \calC, d\rangle$.
 The reduction also gives a one-to-one correspondence $\psi:\calD \to X$ such that
$$\cost_X(\psi(S)) \le \cost_{\calD, d}(S) \le (1+\varepsilon) \cost_{X}(\psi(S)),$$
where $S = \Angle{S_1,\ldots,S_k}$ is a partition of $\calD$ and $\psi(S) = \Angle{\psi(S_1),\ldots,\psi(S_k)}$ is the corresponding partition of $X$.
The reduction runs in time $n^{O(\log(1/\varepsilon) / \varepsilon^2)}$.

2. In  instance $\langle \calD, \calC, d\rangle$, $\calC \subset \bbR^{\tilde p}$ (for some $\tilde p$), $\calC$ is an $(\varepsilon/3)$-approximate centroid for $\calD$, and $d(c,x)= \|c -x\|^2$.
\end{theorem}



\section{Algorithm for $k$-Median with Relaxed Triangle Inequality}
\label{sec:appr-k-medi}
We now turn to the problem of approximating the $k$-median instance from Theorem \ref{thm:reduction-main}.  Our first algorithm is based on
the following standard linear programming relaxation for the $k$-median problem:
\begin{eqnarray}
\min \sum_{c\in {\cal C}}\sum_{x\in X}z_{xc}d(x,c),&&\\
\sum_{c\in  {\cal C}} y_c = k, &&  \label{constr-eq-k}\\
\sum_{c\in {\cal C}} z_{xc}=1, && \forall x\in  {X},\label{constr-eq-1}\\
z_{xc}\le y_{c},&& \forall c\in {\cal C}, j\in  {X},\label{constr-less}\\
 z_{xc},y_c\ge 0.&&\label{constlast}
\end{eqnarray}
In the integral solution, each variable $y_c$ indicates whether the center $c$ is open; and each variable $z_{xc}$ indicates whether
the point $x$ is assigned to the center $c$. Constraint~(\ref{constr-eq-k}) asserts that we should open exactly $k$ centers;
constraint~(\ref{constr-eq-1}) ensures that every point is assigned to exactly one center; finally, constraint~(\ref{constr-less})
says that points can be assigned only to open centers. In a fractional LP solution, all $z_{xc}$ and $y_c$ lie in the interval $[0,1]$.
Note that in the integral solution, $z_{xc}=y_c$, if $z_{xc}>0$ (as both $z_{xc}$ and $y_c$ must be equal to $1$).
We can slightly change any feasible LP solution so it also satisfies this property. Specifically, we split any center
$c$ which does not satisfy $y_c = z_{xc}$ (for some $x\in X$) in two co-located centers $c_1$ and $c_2$: one with weight $z_{xc}$ and the other
with weight $y_c - z_{xc}$. We distribute the weights $z_{x'c}$ among them as follows: we let $z_{x'c_1} = \min(z_{x'c}, y_{c_1})$;
$z_{x'c_2} = y_{c_2} - \min(z_{x'c}, y_{c_1})$. Note that this is a standard assumption in the $k$-median literature.
We refer the reader to~\cite{S} (see Lemma~1) and~\cite{CS} for more details.
The values $y_c$ define the measure $y$ on $\calC$: $y(C)=\sum_{c\in C} y_c$. In the rounding algorithm and in the analysis,
it will be convenient to think of this measure as of ``continuous measure'': That is, if needed we will
split the centers into co-located centers to ensure that we can find a set of any given measure $\mu$.

For every point $x\in  X$, let $C_x = \{c\in \calC: z_{xc}>0\}$. The set $C_x$ contains all centers that serve $x$ in the LP solution.
Recall that we modify the solution so that $y_c=z_{xc}$ if $z_{xc}>0$. Hence, $y_c=z_{xc}$ if $x\in C_x$.
For every point $x\in  X$, we define its LP radius $R_x$ as:
$$R_x = \sum_{c\in \calC} z_{xc} d(x,c) = \sum_{c\in C_x} y_c d(x,c).$$
Observe, that the LP value, which we denote by $LP$, equals $\sum_{x\in X} R_x$.

\textbf{Algorithm.} We now describe our LP-rounding algorithm for $k$-Medians with relaxed $3$-hop triangle inequality.
\begin{theorem}
There exists a $(\beta,\alpha)$ bi-criteria approximation algorithm for $k$-means with 
\begin{equation}\label{eq:LP-alpha}
\alpha(\beta) = 1+e^{-\beta}\Big(\frac{6\beta}{1-\beta} + \frac{(\beta-1)^2}{\beta}\Big)
\end{equation}
for every $\beta > 1$.
\end{theorem}

The algorithm first solves the LP problem and modifies the LP solution as described above if necessary. Then, it partitions all centers into
$\beta k$ groups $Z \in \calZ$, each with LP measure $1/\beta$. It
picks one center $c$ at random  from each group $Z$ with probability $\beta y_{c}$ (note, that $\sum_{c\in Z} \beta y_{c} = 1$).
The algorithm outputs the set of $\beta k$ chosen centers, and assigns every point to the closest center.

We now describe the construction of $\calZ$ in more detail.  We partition centers into $\beta k$ groups as follows.
For every $x\in  X$, we find the unique ball $B_x$ around $x$ whose LP weight exactly
equals $1/\beta$ (To do so, we may split some centers, and pick some centers in $B_x$ at the boundary of the ball but
not the others). We find a subset of points $\calW$ such that balls $B_x$ with $x\in \calW$ are disjoint, and
for every point $x\in  X$, we also define a ``witness'' $w(x)\in \calW$. To this end,
we sort all points $x\in  X$ by the LP radius $R_x$ in the ascending order, and then
consider them one by one. For each $x\in  X$, if $B_x$ is disjoint from all previously chosen balls,
then we add $x$ to the set $\calW$, we set $w(x)=x$. Otherwise, if $B_x$ intersects some other ball
$B_{x'}$ that is already chosen, we discard $B_x$ and set $w(x) = x'$.
If there are several balls $B_{x'}$ intersecting $B_x$, we pick the first $x'$ according to our
ordering as the witness. Note, that $R_{w(x)}\leq R_x$ for all $x$. Once, we found a disjoint
 collection of balls $\{B_x: x\in {\cal W}\}$, we add them to the set $\calZ$. We partition centers
not covered by $\cup_{x\in\calW}B_x$ into groups of LP weight $1/\beta$ arbitrarily
and add these groups to $\calZ$. Thus, we obtain a partitioning $\calZ$ of all centers
into groups of LP weight $1/\beta$.

\textbf{Analysis. }
We show that the algorithm returns a valid solution, and then prove an upper bound on its expected cost.
 The algorithm picks
exactly one vertex from each group, so it always picks $\beta k$ vertices. Hence, it always outputs a valid solution.

We now give an overview of the proof of the upper bound, and then present the details.
Let $S$ be the set of centers output by the algorithm. Denote the radius of the ball $B_x$ by
$R_x^{\beta}$. For every vertex $x$, we estimate the expected distance from $x$
to the closest center $c$ in the solution $S$ i.e. $\E[d(x,S)]$. We show that
$\E[d(x,S)]\leq \alpha(\beta)\,R_x$ for $\alpha(\beta)$ as in equation~(\ref{eq:LP-alpha}). Since
$LP=\sum_x R_x$, we conclude that the algorithm has an approximation factor of $\alpha(\beta)$.

Fix $x\in  X$. Recall, that $C_x = \{c: z_{xc}>0\}$ is the set of all
centers that serve $x$ in the LP solution.
We upper bound $d(x,S)$ by $d(x, (C_x\cup B_{w(x)})\cap S)$, which is
the distance to the closest center in $C_x\cup B_{w(x)}$ chosen by the algorithm. Note that the solution $S$ always
contains at least one center in $B_{w(x)}$, so $(C_x\cup B_{w(x)})\cap S\neq \varnothing$.
For the proof, we pick a particular (random) center $f(x)\in (C_x\cup B_{w(x)})\cap S$.

We define $f(x)$ using the following randomized procedure.
Consider the partitioning $\calZ$ of all centers into groups of measure $1/\beta$ used by the algorithm.
Let $\widetilde\calZ = \{Z\cap C_x: Z\in \calZ;\,Z\cap C_x\neq \varnothing\}$ be the induced partitioning of the set $C_x$. For all $\widetilde{Z}\in \widetilde{\calZ}$ we independently flip a coin and with probability $(1-e^{-\beta y(\widetilde{Z})})/(\beta y(\widetilde{Z}))$
make the set $\widetilde{Z}$ \textit{active}. We let $A\subset C_x$ to be the union of all active sets $\widetilde{Z}$;
we say that centers in $A$ are active centers. Let $f(x)$ be the center in $A\cap S$ closest to $x$, if $A\cap S\neq \varnothing$ ;
let $f(x)$ to be the unique center in $B_{w(x)}\cap S$, otherwise.
We set $\calE = 0$, if $A\cap S\neq \varnothing$; and $\calE = 1$, otherwise.
Roughly speaking, $\calE$ indicates whether $f(x)\in C_x$ or $f(x)\in B_{w(x)}$: Specifically, if $\calE = 0$, then
$f(x)\in C_x$; if $\calE = 1$, then $f(x)\in B_{w(x)}$. Note, however, that $C_x\cap B_{w(x)}\neq \varnothing$,
and $f(x)$ may belong to $C_x\cap B_{w(x)}$.

The center $f(x)$ may not be the closest to $x$, but since $f(x)\in S$, we have
$$d(x,S)\leq d(x, (C_x\cup B_{w(x)})\cap S)\leq d(x, f(x)).$$
In Lemma~\ref{lem:probE}, we show that $\Pr(\calE=0) = 1 - e^{-\beta}$. Thus,
\begin{eqnarray*}
E[d(x, f(x)] &=& \Pr(\calE = 0)\, \E\big[d(x,f(x))\given \calE = 0\big] + \Pr(\calE=1)\, \E\big[d(x,f(x))\given  \calE = 1\big]\\
&=& (1-e^{-\beta})\, \E\big[d(x,f(x))\given \calE = 0\big] +  e^{-\beta}\, \E\big[d(x,f(x))\given \calE = 1\big].
\end{eqnarray*}
We bound the expected distance from $x$ to $f(x)$ given $\calE=0$ in Lemma~\ref{lem:bound-U}. We show that
\begin{equation}
\E[d(x,f(x))\given \calE = 0]\leq R_x.
\end{equation}
Observe that for a random center $c$ distributed according to the LP measure in $C_x$  (i.e., $\Pr (c = c_0) = y(c_0)/y(C_x) = y(c_0)$),
we have the exact equality $\E[d(x,c)] = R_x$. So Lemma~\ref{lem:bound-U} shows that
the distribution of $f(x)$ given $\calE = 0$ is ``not worse'' than the distribution according to $y$ in $C_x$.
We now proceed to bound the expected distance from $x$ to $f(x)$ given $\calE=1$.
Recall, that $w(x)$ is the witness for $x$. Thus, the balls $B_{x}$ and $B_{w(x)}$ intersect
and $R_{w(x)}\leq R_x$. Let $c_{\circ}$ be an arbitrary center in $B_{x}\cap B_{w(x)}$.
By the relaxed 3-hop triangle inequality,
\begin{align*}
d(x,f(x))&\leq 3 \big(d(x,c_{\circ})+ d(w(x), c_{\circ}) + d (w(x), f(x)\big)\\
&\leq 3 \big(R^{\beta}_x+ R^{\beta}_{w(x)} + d (w(x), f(x))\big).
\end{align*}
Here, we used that $R^{\beta}_x$ is the radius of $B_x$; $R^{\beta}_{w(x)}$ is the radius of $B_{w(x)}$.
By the Markov inequality, $R^{\beta}_x\leq \beta R_x /(\beta - 1)$ (see Lemma~\ref{lem:R-by-Markov}).
In Lemma~\ref{lem:bound-BW},we show that there exists two nonnegative numbers $r_1$ and $r_2$ ($r_1\leq r_2$) such that
$$R_{w(x)}^{\beta} + \E[d(w(x),f(x))\given f(x)\in B_{w(x)}\setminus D_x;\, \calE=1]\leq r_1 + r_2,$$
and
\begin{equation}\label{eq:lin-eq-r1-r2}
\Big(\frac{1-\gamma}{\beta}\Big)\, r_1 + \Big(\frac{\beta - 1}{\beta}\Big)\, r_2\leq R_x,
\end{equation}
where $\gamma$ is some parameter in $[0,1]$. Hence,
$$\E[d(x,f(x))\given f(x)\in B_{w(x)}\setminus D_x;\, \calE=1]\leq 3\Big(\frac{\beta R_x}{\beta-1} + r_1+r_2\Big).$$
By Lemma~\ref{lem:prob-DU},
$$\Pr (f(x)\in B_{w(x)}\setminus D_x\given \calE=1) = e^{\gamma} (1-\gamma).$$
Finally, in Lemma~\ref{lem:bound-DU}, we show that
$$\E[d(x,f(x))\given f(x)\in D_x;\, \calE=1]\leq \frac{\beta R_x}{\gamma}.$$
Combining all bounds above we get the following inequality:
$$\E[d(x,f(x))]\leq \Big((1-e^{-\beta}) + e^{-\beta}\big(e^{\gamma} (1-\gamma)\times 3\Big(\frac{\beta}{\beta-1} + \frac{r_1 + r_2}{R_x}\Big) +
(1 - e^{\gamma} (1-\gamma))\times \frac{\beta}{\gamma}\big)\Big)\;R_x.$$
We now find the maximum of the right hand side over all possible values of $\gamma\in[0,1]$ and $r_1,r_2\geq 0$ satisfying linear
inequalities $r_1\leq r_2$ and (\ref{eq:lin-eq-r1-r2}). The right hand side is a linear function of $r_1$ and $r_2$. Hence, for
a fixed $\gamma$ the maximum is attained at one of the two extreme points: $(r_1,r_2) = (0, \beta R_x/(\beta-1))$ or
$(r_1,r_2) = (\beta R_x/(\beta-\gamma), \beta R_x/(\beta-\gamma))$.
Substituting $r_1$ and $r_2$ in the previous inequality we get the following bound on the ratio $\E[d(x,f(x))]/R_x$:
\begin{equation}\label{eq:exact-LP-bound}
\max_{\gamma\in[0,1]}
\Big((1-e^{-\beta}) +3e^{-(\beta-\gamma)} (1-\gamma)\Big(\frac{\beta}{\beta-1} + \max\big(\frac{\beta}{\beta-1},\frac{2\beta}{\beta-\gamma}\big)\Big) +
\frac{\beta e^{-\beta}(1 - e^{\gamma} (1-\gamma))}{\gamma}\Big).
\end{equation}
This function can be upper bounded by $\alpha(\beta)$ defined in~(\ref{eq:LP-alpha}).
We conclude that the approximation factor of the algorithm is upper bounded by $\alpha(\beta)$.

\subsection{Detailed Analysis of the LP Rounding Algorithm}

\begin{lemma}\label{lem:probE}
We have $\Pr(\calE = 0)= 1 - e^{-\beta}$.
\end{lemma}
\begin{proof}
Recall, that the algorithm picks one center $c$ in every $Z\in\calZ$ uniformly (with respect to the measure $y$) at random. Thus, the probability
that the algorithm picks a center from $\tilde{Z}$ equals $\beta y(\tilde{Z})$. The probability that
a given $\tilde{Z}$ contains a point from the solution $S$ and
$\tilde{Z}$ is active equals $\beta y(\tilde{Z})\times (1-e^{\beta y(\widetilde{Z})})/(\beta y(\widetilde{Z})) = (1-e^{\beta y(\widetilde{Z})})$.
The probability that no such $\widetilde{Z}$ exists equals
$$\prod_{\widetilde{Z}\in \widetilde{\calZ}} e^{-\beta y(\widetilde{Z})} =
e^{-\sum_{\widetilde{Z}\in \widetilde{\calZ}} \beta y(\widetilde{Z})} = e^{-y(C_x)}= e^{-\beta}.$$
\end{proof}
We now bound $E[d(x,f(x))\given \calE = 0]$.
\begin{lemma}\label{lem:bound-U}
We have $E[d(x,f(x))\given \calE = 0]\leq R_x$.
\end{lemma}
\begin{proof}
We define two sets of random variables $P$ and $Q$, and then show that they are identically distributed. If the algorithm picks
a center $c$ in $\widetilde{Z}$, and $\widetilde{Z}$ is active, let $P(\widetilde{Z}) = c$. Let $P(\widetilde{Z}) = \perp$, otherwise. The
random variables $P(\widetilde{Z})$ are mutually independent for all $\widetilde{Z}\in \widetilde{\calZ}$; and
$$\Pr (\widetilde{Z} = c) = \frac{(1 - e^{-\beta y(\widetilde{Z})}) \,y_c}{y(\widetilde{Z})}$$
for $c\in \widetilde{Z}$.

To define $Q$, we introduce an auxiliary Poisson arrival process.  At every point of time $t\in[0,\beta]$, we pick a center $c\in C_x$ with probability
$y_{c} dt$ (i.e., with arrival rate $y_{c}$). For every $\widetilde{Z}$, let $Q(\widetilde{Z})$ be the first center chosen in $\widetilde{Z}$.
If no centers in $\widetilde{Z}$ are chosen, we let $Q(\widetilde{Z})=\perp$. Note that we pick two centers at exactly the same time with
probability $0$, hence $Q(\widetilde{Z})$ is well defined. Conditional on $Q(\widetilde{Z})\neq \perp$, the random variable
$Q(\widetilde{Z})$ is uniformly distributed in $\widetilde{Z}$ with respect to LP weights $y_c$ (since at every given time $t$, the probability of arrival equals $y_c dt$).
Then, $\Pr (Q(\widetilde{Z})\neq \perp) = (1-e^{-\beta y (\widetilde{Z})})$. Hence, $\Pr (Q(\widetilde{Z})=c) = (1-e^{-\beta y (\widetilde{Z})})y_c/y(\widetilde{Z})$.
Note that all random variables $Q$ are mutually independent. Thus, the random variables $Q$ have the same distribution as random variables $P$.

Note that if $\calE = 0$, then $f(x)$ is the closest center in $\{P(\widetilde{Z}):\widetilde{Z}\in \widetilde{\calZ};\widetilde{Z}\neq \perp \}$ to $x$. If $\calE = 1$, then all
$P(\widetilde{Z})$ are equal to $\perp$. Let $U_Q=\{Q(\widetilde{Z}):\widetilde{Z}\in \widetilde{\calZ};\widetilde{Q}\neq \perp \}$.
Since $P$ and $Q$ have the same distribution, we have
$$\E[d(x, f(x))\given \calE = 0] = \E[\min_{c\in U_Q} d(x,c) \given U_q\neq \varnothing].$$
Conditional on $U_q\neq \varnothing$, the first center that arrives according to our stochastic process is uniformly distributed in $C_x$.
The expected distance from it to $x$ equals $R_x$. This center belongs to $U_c$. Hence, $\E[\min_{c\in U_Q} d(x,c) \given U_q\neq \varnothing]\leq R_x$.
\end{proof}

Let $D_x = B_{w(x)}\cap C_x$ and $\gamma= \beta y(D_x)$. Note that $\gamma\in[0,1]$, since $y(B_{w(x)})=1/\beta$.
We find $\Pr(f(x) \in D_x \given \calE = 1)$.

\begin{lemma}\label{lem:prob-DU}
We have
$$\Pr(f(x) \in D_x \given \calE = 1) = 1- e^{\gamma} (1-\gamma).$$
\end{lemma}
\begin{proof}
Observe that the set $D_x = B_{w(x)}\cap C_x$ is one of the sets in the partitioning $\widetilde \calZ$ as
$w(x)\in \calW$ and $B_{w(x)}\in \calZ$. Assume $f(x)\in D_x$ and $\calE =1 $. Since $f(x)\in D_x$, we have $S\cap D_x \neq \varnothing$. Thus, $D_x$ must be inactive
(otherwise, $\calE$ would be~$0$). Moreover, for every $\widetilde Z \neq D_x$ ($\widetilde Z\in \calZ$),
$\widetilde Z$ is inactive or $\widetilde Z\cap S=\varnothing$ (again, otherwise, $\calE$ would be~$0$).
Hence, the event
 $\{f(x) \in D_x \text{ and } \calE = 1\}$ can be represented
as the intersection of the following three independent events:
$\{S\cap D_x \neq \varnothing \}$, $\{D_x \text{ is not active}\}$, and
$\{\text{there are no active vertices in } (C_x\setminus D_x)\cap S\}$.
The probability of the first event is $\beta y(D_x)$; the probability of the second event is
$1 - (1 - e^{-\beta y(D_x)})/(\beta y(D_x))$; the probability of the third event is
$e^{-\beta y(C_x\setminus D_x)}$
(this probability is computed as in Lemma~\ref{lem:probE}). Thus,
\begin{align*}
\Pr(f(x) \in D_x \text{ and } \calE = 1) &=
\beta y(D_x)\times \Big(1 - \frac{1 - e^{-\beta y(D_x)}}{\beta y(D_x)}\Big)\times e^{-\beta y(C_x\setminus D_x)}\\
&=
\big(\gamma - (1 - e^{-\gamma})\big)\times e^{-(\beta -\gamma)} = e^{\beta}\big(1 - (1-\gamma)e^{\gamma}\big).
\end{align*}
This finishes the proof.
\end{proof}

We bound $\E[d(x, f(x))\given f(x)\in D_x;\,\calE=1]$ and $\E[d(x, f(x))\given f(x)\in B_{w(x)}\setminus D_x;\,\calE=1]$ in Appendix~\ref{appendix:tech-details-alg-alt}.



\section{Local Search}
\label{sec:local-search}
For smaller values of $\beta$, we consider the standard local search
algorithm (see, e.g., \cite{Arya2004}) for the $\beta k$-median
problem using swaps of size $p$. The
algorithm works as follows: we maintain a current solution $A$
comprising $\beta k$ centers in $\calC$. We repeatedly attempt
to reduce the cost of the current solution $A$ by closing a set of
at most $p$ centers in $A$ and opening the same number of new
centers from $\calC \setminus A$. When no such local swap
improves the cost of the solution $A$ we terminate and return $A$.
In order to simplify our analysis, we do not worry about
convergence time of the algorithm here. We note that by applying
standard techniques (see \cite{Arya2004,Charikar2005}), we can
ensure that, for any $\delta > 0$, the algorithm converges in time polynomial in $n = |\calC \cup \calD|$ and $\frac{1}{\delta}$
by instead stopping when no local swap improves the cost of $A$ by a factor of $\left(1 - \frac{\delta}{\poly(n)}\right)$; the resulting algorithm's approximation ratio increases by only $\frac{1}{1-\delta}$.

Unfortunately the analyses of \cite{Arya2004,Charikar2005} relies
heavily on the triangle inequality, while the instances generated
by Theorem \ref{thm:reduction-main} satisfy only a 3-relaxed 3-hop
triangle inequality. Thus, we proceed as in \cite{Kanungo2004}.

Let $O=\Angle{o_1,\ldots,o_k}$ be an optimal
set of $k$
centers, and $A=\Angle{a_1,\ldots,a_{\beta k}}$ be the set of
$\beta k$ centers produced by the local search algorithm. As in
\cite{Kanungo2004}, we say that a center $a \in A$
\emph{captures} a center $o \in O$ if $a$ is the center of $A$
that is closest to $o$. Note that each center in $A$ can
potentially capture several centers in $O$, but each center in
$O$ is captured by exactly one center of $A$. We now construct a
set of local swaps to consider in our analysis. We say that a
center in $A$ is ``good'' if it does not capture any center of
$O$. Then, because each center of $O$ is captured by only one
center of $A$, we must have at least $\beta k - k = (\beta - 1)k$
good centers in $A$. We fix some such set of $(\beta - 1)k$ good
centers; we call them ``auxiliary'' centers and set them
aside for now.

For the remaining $k$ centers $B \subseteq A$, we proceed
exactly as in \cite{Kanungo2004}: we assign each center in $O$ to
the bad center of $B$ that captures it. This creates a partition
$O_1,\ldots,O_r$ of centers in $O$. We similarly partition the
centers of $B$ into $r$ parts $B_1,\ldots,B_r$ with $|B_i| =
|O_i|$; for each $1 \le i \le r$, let $B_i$ contain the bad
center of $B$ that captures all of $O_i$ together with $|B_i| -
1$ unique good centers of $B$. Note that the fact that each
center of $O$ is captured only once ensures that there are indeed
enough good centers in $B$ for our construction. Now, we use
this partition of $B$ and $O$ to construct a set of swaps, each
assigned some weight. If $|O_i| \le p$, we consider the $\langle
B_i, O_i \rangle$ with weight 1. If $|O_i| = q > p$, we consider a
group of singleton swaps $\langle \{b\}, \{o\}\rangle$, where
$o\in O_i$ and $b$ is a good center in $B_i$, each with weight $\frac{1}{q-1}$.
At this
point, note that every center in $O$ occurs in swaps of total
weight $1$, and every center in $B$ occurs in swaps of total
weight at most $\frac{q}{q-1} \le 1 + \frac{1}{p}$. Now, we add swaps involving auxiliary centers; for
each of the $(\beta - 1)k$ auxiliary centers $a \in A \setminus
B$ and each $o \in O$, we consider singleton swap $\langle \{a\},
\{o\} \rangle$, assigned weight $\frac{1}{k}$. Each center of $O$
now occurs in swaps of total weight $1 + (\beta - 1) = \beta$,
while each center of $A \setminus B$ occurs in swaps of total
weight $1$.

Summarizing, our set of swaps satisfies the following properties:
(1) each center of $O$ occurs in swaps of total weight $\beta$;
(2) each center of $A$ occurs in swaps of total weight at most $1
+ \frac{1}{p}$; (3) for any swap $\langle A',O' \rangle$ in our
set, no center in $A'$ captures any center not in $O'$. We now
give a brief sketch of how these properties lead to our desired
approximation ratio (we give a full description of the analysis in
the appendix). Our analysis closely follows that of
\cite{Kanungo2004}.

As in \cite{Kanungo2004}, the total change $\cost_{\calD, d}(A
\setminus A' \cup O') - \cost_{\calD, d}(A)$ due to performing a
single swap $\langle A', O'\rangle$ is at most:
\begin{equation*}
\sum_{o \in O'}\sum_{x \in N_O(o)}\big(d(x,o) - d(x,a_x)\big) +
\sum_{a \in A'}\sum_{x \in N_A(A')}\big(d(x,a_{o_x}) - d(x, a_x)\big).
\end{equation*}
If $A$ is locally optimal, then we must have that $\cost_{\calD,
d}(A \setminus A' \cup O') - \cost_{\calD,d}(A) \ge 0$ for all
swaps $(A',O')$ considered by the algorithm. In particular, for
each swap $\langle A', O'\rangle$ in our set, we have:
\begin{equation}
\label{eq:local-eq}
0 \le \sum_{o \in O'}\sum_{x \in N_O(o)}\big(d(x,o) - d(x,a_x)\big) +
\sum_{a \in A'}\sum_{x \in N_A(A')}\big(d(x,a_{o_x}) - d(x, a_x)\big).
\end{equation}
Multiplying each inequality \eqref{eq:local-eq} by the weight of
its swap and then adding the resulting inequalities we obtain:
\begin{equation*}
\label{eq:global-eq}
0 \le \beta \sum_{x \in \calD}(d(x, o_x) - d(x, a_x)) + \left(1 +
\frac{1}{p}\right)\sum_{x \in \calD}(d(x, a_{o_x}) - d(x, a_x)),
\end{equation*}
due to properties (1) and (2) of our set of swaps. Theorem
\ref{thm:reduction-main} part~2, which shows that
our center set is an approximate $k$-means centroid set, then
allows us to simplify the final term above as in
\cite{Kanungo2004}, giving:
\begin{equation*}
0 \le \left(\beta + {\textstyle 2 + \frac{2}{p}}\right)\cost_{\calD,d}(O) - \left( \beta -
\frac{2 + \frac{2}{p}}{\alpha}\right)\cost_{\calD, d}(A) +
O(\epsilon)\cdot\cost_{\calD,d}(A),
\end{equation*}
where $\alpha^2 = \frac{\cost_{\calD,d}(A)}{\cost_{\calD,d}(O)}$ is
the squared approximation ratio of our algorithm. Rearranging and
simplifying (again, we give a detailed analysis in the appendix),
we obtain.
\[
\alpha < \left(1 + \frac{2}{\beta} + \frac{2}{\beta p}\right)\frac{1}{1 - O(\epsilon)}
\]
Therefore, we have proved the following theorem:
\begin{theorem}
There exists an algorithm that produces a solution for any instance
of $\beta k$-median problem satisfying the properties of Theorem
\ref{thm:reduction-main}, where $\beta>1$ is a fixed constant.  For any $p \ge 1$ and any $\varepsilon \in (0,1]$, the
algorithm runs in time polynomial in $|\calC \cup \calD|$ and
produces a solution $A$ satisfying:
\[
\cost_{\calD, d}(A) \le \left(1 + \frac{2}{\beta} + \frac{2}{\beta p}\right)^2\frac{1}{1 - O(\epsilon)}\cdot\cost_{\calD, d}(O)
\]
where $O$ is the optimal set of $k$ centers in $\cal C$.
\end{theorem}



\bibliographystyle{plain}
\bibliography{colt}

\appendix


\section{Detailed Analysis of the LP Rounding Algorithm}\label{appendix:tech-details-alg-alt}
In this section, we bound $\E[d(x, f(x))\given f(x)\in D_x;\,\calE=1]$ and $\E[d(x, f(x))\given f(x)\in B_{w(x)}\setminus D_x;\,\calE=1]$.

\begin{lemma}\label{lem:bound-DU}
The following bound holds:
$$\E[d(x, f(x))\given f(x)\in D_x;\,\calE=1]\leq \frac{\beta R_x}{\gamma}.$$
\end{lemma}
\begin{proof}
Given $f(x)\in D_x$ and $\calE=1$, the random center $f(x)$ is distributed uniformly in $D_x$ (with respect to the LP weights $y$).
Hence, $\Pr (f(x)=c) = y_c/y(D_x)$ for $c\in D_x$. We have
$$\E[d(x, f(x))\given f(x)\in D_x;\,\calE=1] = \frac{\sum_{c\in D_x} y_c d(x,c)}{y(D_x)}\leq \frac{\sum_{c\in C_x} y_c d(x,c)}{y(D_x)} = \frac{R_x}{\gamma/\beta}.$$
\end{proof}

\begin{lemma}\label{lem:bound-BW}
There exists two nonnegative numbers $r_1$ and $r_2$ satisfying
\begin{enumerate}
\item $$\Big(\frac{1-\gamma}{\beta}\Big)\, r_1 + \Big(\frac{\beta - 1}{\beta}\Big)\, r_2\leq R_x,$$
\item $$r_1\leq r_2,$$
\end{enumerate}
such that
$$R_{w(x)}^{\beta} + \E[d(w(x),f(x))\given f(x)\in B_{w(x)}\setminus D_x;\, \calE=1]\leq r_1 + r_2.$$
\end{lemma}
\begin{proof}
Denote the expected distance from a random center $c$ in $B_{w(x)}\setminus D_x$ to $w(x)$ by $r_1$
and  distance from a random center $c$ in $C_{w(x)}\setminus B_{w(x)}$ to $w(x)$ by $r_2$:
$$r_1 = \sum_{c\in B_{w(x)}\setminus D_x} y_c d(w(x),c);\;\;\;
r_2 = \sum_{c\in C_{w(x)}\setminus B_{w(x)}} y_c d(w(x),c).
$$
By the definition of $R_{w(x)}$, we have
\begin{align*}
R_{w(x)} &= \Big(\sum_{c\in D_x} y_c d(w(x),c)\Big) +
y(B_{w(x)}\setminus D_x)\, r_1 + y(C_x\setminus B_x)\, r_2\\
&\geq \Big(\frac{1-\gamma}{\beta}\Big)\, r_1 + \Big(\frac{\beta - 1}{\beta}\Big)\, r_2.
\end{align*}
Note that $R_{w(x)}\leq R_x$. Hence,
$$\Big(\frac{1-\gamma}{\beta}\Big)\, r_1 + \Big(\frac{\beta - 1}{\beta}\Big)\, r_2\leq R_x.$$
Since all centers in $B_{w(x)}\setminus D_x$ lie inside of the ball of radius $R^{\beta}_{w(x)}$
around $w(x)$, and all centers in $C_{w(x)}\setminus B_{w(x)}$ lie outside of this ball, we have
$r_1\leq R^{\beta}_{w(x)}\leq r_2$. Hence, $R_{w(x)}+d(w(x),f(x))\leq r_2  +d(w(x),f(x))$ and
$r_1\leq r_2$. Conditional on $f(x)\in B_{w(x)}\setminus D_x$ and $\calE=1$, the random
center $f(x)$ is distributed uniformly in $B_{w(x)}\setminus D_x$ with respect to the weights $y$. Hence,
$\E[d(w(x),f(x))\given f(x)\in B_{w(x)}\setminus D_x;\, \calE=1] = r_1$. Consequently,
$$R_{w(x)}^{\beta} + \E[d(w(x),f(x))\given f(x)\in B_{w(x)}\setminus D_x ;\, \calE=1]\leq r_1 + r_2.$$
\end{proof}

\begin{lemma}\label{lem:R-by-Markov}
The following inequality holds:
$R_x^{\beta}\leq \beta R_x/(\beta -1)$.
\end{lemma}
\begin{proof}
We have
$$R_x = \sum_{c\in C_x} y_c d(x,c) \leq \sum_{c\in C_x\setminus B_x} y_c d(x,c).$$
Every center $c\in C_x\setminus B_x$ is at distance at least $R^{\beta}_x$ from $x$. Hence,
$$R_x \leq \sum_{c\in C_x\setminus B_x} y_c R^{\beta}_x = y(C_x\setminus B_x)\, R^{\beta}_x= \big(1-\frac{1}{\beta}\big)R^{\beta}_x.$$
The desired inequality follows.
\end{proof}



\section{Detailed Analysis of the Local Search Algorithm}
\label{sec:deta-analys-local}

Here we give a detailed analysis of the local search algorithm from section \ref{sec:local-search}, closely following \cite{Kanungo2004}.

For a set of points $P \subseteq X$ and a point $c \in \bbR^p$, define the total distortion of $P$ with respect to $c$ as
$\Delta(P,c) \equiv \sum_{p \in P}\|p - c\|^2$.  We shall use the following Lemmas from \cite{Kanungo2004}:
\begin{lemma}[{Lemma 2.1 in \cite{Kanungo2004}}]\label{lem:kanungo}
Given a finite subset $P$ of points in $\bbR^p$, let $c$ be the centroid of $P$.  Then, for any $c' \in \bbR^p$, $\Delta(P,c') = \Delta(P,c) + |P|\cdot \|c - c'\|^2$
\end{lemma}

\begin{lemma}\label{lem:kanungo2}
Let $\langle \rho_i \rangle$ and $\langle \xi_i \rangle$ be two sequences of reals such that $\alpha^2 = (\sum_i \rho_i^2)/(\sum_i\xi_i^2)$ for some $\alpha > 0$.  Then,
\[
\sum_{i = 1}^n \rho_i\xi_i \le \frac{1}{\alpha}\sum_{i = 1}^n \rho_i^2.
\]
\end{lemma}

We now show how local optimality implies the desired inequality.  For a demand point $x \in \calD$, let $a_x$ and $o_x$ denote the closest facility to $x$ in $A$ and $O$, respectively. Recall that for for $a \in A$, $N_A(a)$ is precisely the set of all those demand points $x \in \calD$ such that $a_x = a$, and, similarly, for $o \in O$, $N_O(o)$ is the set of all demand point $x \in \calD$ such that $o_x = o$.  Now, we upper bound the change in cost due to some swap $\langle A', O' \rangle$ in our set of swaps.  We do this by constructing a feasible assignment of all points in $\calD$ to centers in $A \setminus A' \cup O'$.  For each $o \in O'$, we assign all the points in $N_O(o)$ to $o$.  This changes the cost by
\[
\sum_{o \in O'}\sum_{x \in N_O(o)}(d(x,o) - d(x,a_x)).
\]
Now, fix a point $x \in N_A(A') \setminus N_O(O')$, and consider $x$'s closest optimal
center $o_x$.  We must have $o_x \not\in O'$.  Let $a_{o_x}$ be the closest center to $o_x$ in $A$.  Then, by property (3) above, $a_{o_x} \not\in A'$, since $a_{o_x}$ captures $o_x$ but $o_x \not\in O'$.  We reassign $x$ to $a_{o_x}$.  The total cost of reassigning all such points $x$ is at most:
\[
\sum_{a \in A'}\sum_{x \in N_A(A') \setminus N_O(O')}(d(x,a_{o_x}) - d(x, a_x)) \le
\sum_{a \in A'}\sum_{x \in N_A(A')}(d(x,a_{o_x}) - d(x, a_x)),
\]
where the inequality follows from the fact that $a_x$ is the closest center to $x$ in $A$, and so $d(x,a_{o_x}) - d(x,a_x) \ge 0$ for all $x \in N_A(A') \cap N_O(O')$.  Thus, the total change $\cost_{\calD, d}(A \setminus A' \cup O') - \cost_{\calD,d}(A)$ for each swap $\langle A', O' \rangle$ is at most:
\begin{equation*}
\sum_{o \in O'}\sum_{x \in N_O(o)}(d(x,o) - d(x,a_x) + \sum_{a \in A'}\sum_{x \in N_A(A')}(d(x,a_{o_x}) - d(x, a_x)).
\end{equation*}

If $A$ is locally optimal, then we must have that $\cost_{\calD, d}(A \setminus A' \cup O') - \cost_{\calD,d}(A) \ge 0$ for all swaps $(A',O')$ considered by the algorithm.  In particular, for each swap $\langle A', O'\rangle$ in our set, we have:
\begin{equation*}
0 \le \sum_{o \in O'}\sum_{x \in N_O(o)}\big(d(x,o) - d(x,a_x)\big) + \sum_{a \in A'}\sum_{x \in N_A(A')}\big(d(x,a_{o_x}) - d(x, a_x)\big).
\end{equation*}
Set $\gamma = 1 + \frac{1}{p}$.  Then, multiplying each such inequality by the weight of its swap and then adding the resulting inequalities we obtain
\begin{align}
0 &\le \beta \sum_{x \in \calD}\big(d(x, o_x) - d(x, a_x)\big) + \gamma\sum_{x \in \calD}\big(d(x, a_{o_x}) - d(x, a_x)\big) \notag \\
&= \beta \cost_{\calD, d}(O) - (\beta + \gamma)\cost_{\calD, d}(A) + \gamma\sum_{x \in \calD}d(x,a_{o_x}),\label{eq:allswaps}
\end{align}
where we have exploited properties (1) and (2) of our set of swaps to bound the number of times a given center in $O$ or $A$ is counted in our sum of inequalities.

It remains to bound the final term in \eqref{eq:allswaps}.  Consider some $o \in O$, and let $c$ be the centroid of $N_O(o)$.  As above, we will let $a_o$ denote the closest center in $A$ to $O$.  Then, note that:
\begin{align*}
\Delta(N_O(o), a_{o})
&= \Delta(N_O(o), c) + |N_O(o)| \cdot \|c - a_{o}\|^2 \tag{Lemma \ref{lem:kanungo}} \\
&\le \Delta(N_O(o), o) + |N_O(o)| \cdot \|c - a_{o}\|^2 \tag{$c$ is the centroid of $N_O(o)$}\\
&\le
\sum_{x \in N_O(o)}\left[d(x,o) + (1 + \varepsilon)\|o - a_{o}\|^2\right] \tag{Theorem \ref{thm:reduction-main} part~2, and the fact that $o$ is optimal center for $N_O(o)$} \\
&\le
\sum_{x \in N_O(o)}\left[d(x,o) + (1 + \varepsilon)\|o - a_{x}\|^2\right]. \tag{$a_{o}$ is the closest center to $o$ in $A$} \\
&\le
(1 + \varepsilon)\sum_{x \in N_O(o)}\left[d(x,o) + \|o - a_{x}\|^2\right].
\end{align*}
Let $\alpha^2 = \frac{\cost_{\calD, d}(A)}{\cost_{\calD, d}(O)}=\frac{\sum_{x \in \calD}d(x, a_x)}{\sum_{x \in \calD}d(x, o_x)}$ be the approximation ratio attained by the algorithm.  Summing over all $o \in O$, and recalling that for all $x \in N_O(o)$ we have $o_x = o$, we obtain:
\begin{align}
\sum_{x \in \calD}d(x, a_{o_x}) &= \sum_{o \in O}\Delta(N_O(o), a_{o}) \notag \\
&\le (1 + \varepsilon)\sum_{o \in O}\sum_{x \in N_O(o)}\left[d(x,o) + \|o - a_{x}\|^2\right] \notag \\
&= (1 + \varepsilon)\sum_{x \in \calD}\left[d(x,o_x) + \|o_x - a_{x}\|^2\right] \notag \\
&= (1 + \varepsilon)\sum_{x \in \calD}\left[d(x,o_x) + \|x - o_x\|^2 + \|x - a_x\|^2 + 2\|x - o_{x}\|\|x - a_x\|\right] \notag \\
&= (1 + \varepsilon)\sum_{x \in \calD}\left[2d(x,o_x) + d(x,a_x) + 2\|x - o_{x}\|\|x - a_x\|\right] \notag \\
&\le (1 + \varepsilon)\sum_{x \in \calD}\left[2d(x,o_x) + d(x,a_x) + \frac{2}{\alpha}d(x,a_x)\right] \notag \\
&= (1 + \varepsilon)\left[2\cost_{\calD, d}(O) + \left(1 + \frac{2}{\alpha}\right)\cost_{\calD,d}(A)\right]. \label{eq:Rupperbound}
\end{align}
Where in the last inequality, we have applied Lemma \ref{lem:kanungo2} to the sequences $\rho_i$ and $\xi_i$ defined by: \[\alpha^2 = \frac{\sum_{x \in \calD}d(x, a_x)}{\sum_{x \in \calD}d(x, o_x)} = \frac{\sum_{i = 1}^n \rho_i}{\sum_{i = 1}^n \xi_i}.\]

Applying the upper bound \eqref{eq:Rupperbound} to the final term of \eqref{eq:allswaps}, we obtain:
\begin{align*}
0 &\le \beta \cost_{\calD, d}(O) - (\textstyle \beta + \gamma)\cost_{\calD, d}(A) + \gamma(1 + \varepsilon)\left[2\cost_{\calD, d}(O) + \left(1 + \frac{2}{\alpha}\right)\cost_{\calD,d}(A)\right] \\
&\le (\beta + 2\gamma)\cost_{\calD,d}(O) - \left(\textstyle\beta - \frac{2\gamma}{\alpha}\right)\cost_{\calD, d}(A) + \left(\textstyle3 + \frac{2}{\alpha}\right)\gamma\varepsilon\cost_{\calD,d}(A)
\end{align*}
where we have used the fact that $\cost_{\calD,d}(O) \le \cost_{\calD,d}(A)$.
Rearranging, we have
\begin{align*}
\left(\beta + 2\gamma\right)\cost_{\calD, d}(O) &\ge \left(\beta - \frac{2\gamma}{\alpha} - \left(\textstyle 3+\frac{2}{\alpha} \right)\gamma\epsilon\right)\cost_{\calD,d}(A) \\
&= \left(\beta - \frac{2\gamma}{\alpha} - \left(\textstyle 3+\frac{2}{\alpha} \right)\gamma\epsilon\right)\alpha^2\cost_{\calD,d}(O),
\end{align*}
which implies:
\begin{align*}
\alpha^2\beta - 2\gamma\alpha - \beta - 2\gamma - \alpha^2\left({\textstyle 3 + \frac{2}{\alpha}}\right)\gamma\epsilon &\le 0 \\
\alpha^2 - \frac{2\gamma\alpha}{\beta} - 1 - \frac{2\gamma}{\beta} - \frac{\alpha^2}{\beta}\left({\textstyle 3 + \frac{2}{\alpha}}\right)\gamma\epsilon  &\le 0 \\
(\alpha + 1)\left(\alpha - 1 - \frac{2\gamma}{\beta}\right) -
\frac{\alpha^2}{\beta}\left({\textstyle 3 + \frac{2}{\alpha}}\right)\gamma\epsilon & \le 0 \\
\left(\alpha - 1 - \frac{2\gamma}{\beta}\right) - \frac{\alpha^2}{(\alpha + 1)\beta}\left({\textstyle 3 + \frac{2}{\alpha}}\right)\gamma\epsilon& \le 0.
\end{align*}
Thus, we have:
\begin{align*}
(1 - O(\epsilon)) \alpha &\le 1 + \frac{2\gamma}{\beta} = 1 + \frac{2}{\beta} + \frac{2}{\beta p}.
\end{align*}


\section{Proof of Theorem~\ref{thm:reduction-main}}\label{sec:reduction-proofs}
In this section, we prove Theorem~\ref{thm:reduction-main}.
Consider an instance of $k$-means with a set of points $X\subset \bbR^p$. Denote $n = |X|$. Let $\varepsilon' = \varepsilon / 3$. Let $\varphi:{\mathbb R}^p \to {\mathbb R}^{\tilde p}$ be a dimension reduction transform for $X$ with distortion $(1 + \varepsilon')$ as in Theorem~\ref{thm:JL}.
Note that $\tilde{p} = O(\log n/\varepsilon'^2) = O(\log n/\varepsilon^2)$.

Let $X' = \varphi(X) \subset \bbR^{\tilde{p}}$.
Using the algorithm from Theorem~\ref{thm:matousek}, we compute an $\varepsilon'$-approximate centroid set $\calC \subset \bbR^{\tilde p}$ for $X'$.
The size of $\calC$ is
 $$O(n \varepsilon^{-\tilde{p}}\log (1/\varepsilon)) =
 n \varepsilon^{-O(\log n/\varepsilon^2)}\log (1/\varepsilon) =
 n \cdot n^{O(\log(1/\varepsilon) / \varepsilon^2)} =   n^{O(\log(1/\varepsilon) / \varepsilon^2)};$$
 we need time
$O(n \log n + n \varepsilon^{-\tilde p}\log(1/\varepsilon)) = n^{O(\log(1/\varepsilon) / \varepsilon^2)}$ to compute it.

We first show that for every solution of the $k$-means problem on $X$ there is a corresponding
solution of $k$-means problem on $X'$ in which all centers lie in $\calC$, and vice versa.

\begin{lemma}\label{lem:reduction}
\begin{enumerate}
\item\label{item:red1} For every partition $S=\Angle{S_1 ,\ldots, S_k}$ of $X$, there is a corresponding clustering of $X'$ given by $S'=\Angle{\varphi(S_1), \ldots,\varphi(S_k)}$ and some centers $C' = \Angle{c_1',\ldots,c_k'} \subseteq \calC$ such that:
$$\cost_{X'}(S',C') \le (1+\varepsilon')^2 \cost_X(S).$$
\item\label{item:red2} For every partition $S'=\Angle{S_1',\ldots, S_k'}$ of $X'$, there is a corresponding clustering $S=\Angle{\varphi^{-1}(S_1),\ldots,\varphi^{-1}(S_k)}$ of $X$ and some centers $C = \Angle{c_1,\ldots,c_k} \subseteq \bbR^{p}$ such that $$\cost_{X}(S,C) \le \cost_{X'}(S').$$
\end{enumerate}
\end{lemma}
\begin{proof}
Part \ref{item:red1}: Consider a partition $S=\Angle{S_1, \ldots,  S_k}$ of $X$ and the corresponding partition
$S'=\Angle{S_1', \dots, S_k'}$ of $X'$, where $S_i' = \varphi(S_i)$.  Let $c_i' = \argmin_{c\in \calC} \sum_{x\in S_i'}\|x'-c\|^2$ for $i\in\Set{1,\dots, k}$.  Because $\calC$ is an $\varepsilon'$-approximate centroid set for $X'$, we have, for each cluster $S_i'$,
\begin{align*}\sum_{x\in S_i'}\|x-c_i'\|^2 & \leq (1 + \varepsilon') \min_{c\in \bbR^{\tilde p}}\sum_{x\in S_i'}\|x-c\|^2 =
(1 + \varepsilon') \frac{1}{2|S_i'|} \sum_{x',x''\in S_i'}\|x'-x''\|^2 \\
{}&=  \frac{1 + \varepsilon'}{2|S_i'|}\sum_{x',x''\in S_i}\|\varphi(x')-\varphi(x'')\|^2
\leq  \frac{(1 + \varepsilon')^2}{2|S_i|} \sum_{x',x''\in S_i}\|x'-x''\|^2
\end{align*}
Hence,
$$\cost_{X'}(S') = \sum_{i=1}^k\sum_{x\in S_i'}\|x-c_i'\|^2 \le (1+\varepsilon')^2 \cost_X(S).$$

Part \ref{item:red2}: Consider a partition $S'=\Angle{S_1', \ldots, S_k'}$ of $X'$ and the corresponding partition
$S=\Angle{S_1, \ldots, S_k}$ of $X$, where $S_i = \varphi^{-1}(S_i)$.  Define the centers $c_i = \sum_{x\in S_i} x / |S_i|$.  Then, for each cluster $S_i$, we have:
\begin{align*}
\sum_{x\in S_i}\|x-c_i\|^2 &= \frac{1}{2|S_i|} \sum_{x',x''\in S_i}\|x'-x''\|^2 \leq \frac{1}{2|S_i|} \sum_{x',x''\in S_i}\|\varphi(x')-\varphi(x'')\|^2\\
&=  \frac{1}{2|S_i'|} \sum_{x',x''\in S_i'}\|x'-x''\|^2.
 \end{align*}
Hence,
$$\cost_{X}(S) = \sum_{i=1}^k \sum_{x\in S_i}\|x-c_i\|^2 \leq \cost_{X'}(S').$$
\end{proof}

Now we are ready to define instance $\langle \calD, \calC, d\rangle$.
Let $\calD = X'$, $\calC$ be the $\varepsilon$-approximate centroid we defined above, and $d(c,x) = \|c - x\|^2$ for every $c\in\calC$ and $x\in \calD$.
Define  $\psi:\calD\to X$ by $\psi(x) = \varphi^{-1}(x)$.

We prove that our reduction, which maps instance $X$ of $k$-means to instance  $\langle \calD, \calC, d\rangle$ of $k$-median, satisfies the conditions of the theorem.

\begin{lemma}
 Our reduction produces an instance that satisfies the following properties:
\begin{enumerate}
\item\label{item:main2} The distance function $d$ satisfies the 3-relaxed 3-hop triangle inequality on $\calD \cup \calC$.

\item\label{item:main3} For every partition $S = \Angle{S_1,\ldots,S_k}$ of $\calD$ and the corresponding partition
$\psi(S) = \Angle{\psi(S_1),\ldots,\psi(S_k)}$ of $X$, we have
$$\cost_X(\psi(S)) \le \cost_{\calD, d}(S) \le (1+\varepsilon) \cost_{X}(\psi(S)).$$
\item\label{item:main4} We have
$$\mathrm{OPT}_X \leq \mathrm{OPT}_{\langle \calD, d\rangle} \leq (1+\varepsilon) \mathrm{OPT}_X.$$
\end{enumerate}
\end{lemma}
\begin{proof}

Claim \ref{item:main2} follows from the fact that:
$$\|x - w\|^2 \leq  (\|x - y\|+ \|y - z\| + \|z - w\|)^2 \leq 3 (\|x - y\|^2+ \|y - z\|^2 + \|z - w\|^2).
$$
for any $w,x,y,z \in \bbR^{\tilde{p}}$.

For claim \ref{item:main3}, consider any partition $S$ of $\calD$. Let $T= \psi(S)$ be the corresponding partition of $X$, given by $T_i = \psi(S_i)$.  Then, from our definition of $d$, we have $\cost_{\calD,d}(S) = \cost_{X'}(S)$.   Moreover, by Lemma \ref{lem:reduction}, we have $\cost_{X'}(S)$ is between $\cost_{X}(T)$ and $(1 + \varepsilon')^2 \cost_{X}(T)$.  Thus,
$$
\cost_X(\psi(S)) \le \cost_{\calD,d}(S) \le (1 + \varepsilon')^2\cost_X(\psi(S)) \leq (1 + \varepsilon)\cost_X(\psi(S)).
$$
Since for every partition $S$ of $\calC$ there is a corresponding partition $\psi(S)$ of $X$, and for every partition $T$ of $X$ there is a corresponding partition $\varphi(T)$ of $\calD$, we immediately get from claim \ref{item:main3}
that
$\mathrm{OPT}_X \leq \mathrm{OPT}_{\langle \calD,  d\rangle} \leq (1+\varepsilon) \mathrm{OPT}_X$.

\end{proof}

\end{document}